\documentclass[]{spie}  

\usepackage{amsmath,amsfonts,amssymb}
\usepackage{graphicx}
\usepackage{caption}

\usepackage[colorlinks=true, allcolors=blue]{hyperref}

\title{Recent and Upcoming Upgrades for MIRC-X and MYSTIC on the CHARA Array}

\author[a]{Noura Ibrahim}
\author[a]{Mayra Gutierrez}
\author[a]{John D. Monnier}
\author[b]{Stefan Kraus}
\author[c]{Jean-Baptiste Le Bouquin}
\author[d]{Narsireddy Anugu}
\author[d]{Theo ten Brummelaar}
\author[b]{Sorabh Chhabra}
\author[b]{Isabelle Codron}
\author[e]{Julien Dejonghe}
\author[f]{Aaron Labdon}
\author[e]{Daniel Lecron}
\author[g]{Daniel Mortimer}
\author[e]{Denis Mourard}
\author[d]{Gail Schaefer}
\author[g]{Benjamin Setterholm}
\author[h]{Manuela Arnó}
\author[h]{Andrea Bianco}
\author[h]{Michele Frangiamore}
\author[c]{Laurent Jocou}

\affil[a]{Astronomy Department, University of Michigan, Ann Arbor, MI 48109, USA}
\affil[b]{School of Physics and Astronomy, University of Exeter, Exeter, Stocker Road, EX4 4QL, UK}
\affil[c]{Institut de Planétologie et d'Astrophysique de Grenoble, Grenoble 38058, France}
\affil[d]{The CHARA Array of Georgia State University, Mount Wilson Observatory, Mount Wilson,
CA 91203, USA}

\affil[e]{Laboratoire Lagrange, Université Côte d’Azur, Observatoire de la Côte d’Azur, CNRS, Boulevard de l’Observatoire, CS 34229, 06304 Nice Cedex 4, France}
\affil[f]{European Southern Observatory, Alonso de Cordoba 3107, Vitacura, Santiago, Chile}
 \affil[g]{Max-Planck-Institut für Astronomie, Königstuhl 17, D-69117 Heidelberg, Germany}
\affil[h]{INAF – Osservatorio Astronomico di Brera, Via Bianchi 46, 23807 Merate, Italy}

\authorinfo{Further author information: (Send correspondence to N.I.)\\N.I.: E-mail: inoura@umich.edu}

\begin{document} 
\maketitle

\begin{abstract}
MIRC-X and MYSTIC are six-telescope near-infrared beam (1.08-2.38 $\mu$m) combiners at the CHARA Array on Mt Wilson CA, USA. Ever since the commissioning of MIRC-X (J and H bands) in 2018 and MYSTIC (K bands) in 2021, they have been the most popular and over-subscribed instruments at the array. Observers have been able to image stellar objects with sensitivity down to 8.1 mag in H and 7.8 mag in K-band under the very best conditions. In 2022 MYSTIC was upgraded with a new ABCD mode using the VLTI/GRAVITY 4-beam integrated optics chip, with the goal of improving the sensitivity and calibration. The ABCD mode has been used to observe more than 20 T Tauri stars; however, the data pipeline is still being developed. Alongside software upgrades, we detail planned upgrades to both instruments in this paper. The main upgrades are: 1) Adding a motorized filter wheel to MIRC-X along with new high spectral resolution modes 2) Updating MIRC-X optics to allow for simultaneous 6T J+H observations 3) Removing the warm window between the spectrograph and the warm optics in MYSTIC 4) Adding a 6T ABCD mode to MIRC-X in collaboration with CHARA/SPICA 5) Updating the MIRC-X CRED-ONE camera funded by Prof. Kraus from U. Exeter 6) Carrying out science verification of the MIRC-X polarization mode 7) Developing new software for ABCD-mode data reduction and more efficient calibration routines. We expect these upgrades to not only improve the observing experience, but also increase the sensitivity by 0.4 mag in J+H-bands, and 1 mag in K-band. 
\end{abstract}


\section{INTRODUCTION}
\label{sec:intro}  

The Center for High Angular Resolution Astronomy (CHARA) Array is the largest operating optical/near-infrared interferometer in the world\cite{chara1,chara2}. Based on Mt. Wilson, CA, USA, the array consists of 6 telescopes arranged in a Y formation with a maximum resolving power equivalent to a 330-meter telescope. Only a few instruments at the array combine the light from all six telescopes simultaneously, and here we discuss the ones operating in the near-infrared. The Michigan InfraRed Combiner (MIRC)\cite{mirc1,mirc2} instrument was commissioned in 2005 and was designed to combine the light from all six telescopes, providing simultaneous fringe measurement on all 15 baseline combinations and allowing for a much expanded u,v-plane coverage. While it was operational, MIRC led to many astounding scientific results including imaging spotted stellar surfaces\cite{spots}, the transit of eclipsing binary systems\cite{binary}, disks around Be stars, and the expansion phase of a nova explosion\cite{nova}. 

In June of 2017, the combiner underwent the first phase of upgrades, which replaced the old PICNIC camera with a more sensitive one, funded by Prof. Kraus from U. Exeter. The C-RED ONE \cite{cred} camera features sub-electron readout noise and fast frame rate (3500Hz) with state-of-the-art HgCdTe Electron Avalanche Photodiode Array (eAPD) technology \cite{cred2}. The second upgrade phase included optimizing the beam train optics, which concluded the commissioning of the new MIRC-X instrument in 2018 \cite{mircx1,mircx2}. The upgrade improved the sensitivity by $\sim 2$ magnitudes in H-band (1.65µm), allowing for the imaging of faint and extended objects such as the inner regions of disks around young stellar objects (YSOs)\cite{yso1,yso2,yso3}. 

A few years later, the Michigan Young STar Imager at CHARA (MYSTIC) was commissioned in 2021\cite{mystic}, operating in K-band (2.2µm). This cryogenic sister instrument was designed by the same team alongside MIRC-X and is equipped with a similar C-RED ONE camera. The two instruments are co-phased for simultaneous multiwavelength observing and fringe tracking. 

This paper aims to describe recent and upcoming upgrades to both instruments. The upgrades will provide new observing modes for users as well as new high-resolution spectral modes. In addition, we detail automation plans to provide a smoother and more efficient user experience. 

\section{Instrument Upgrades} 
\subsection{New Observing Modes}
\label{sec:title}
\subsubsection{MYSTIC ABCD Mode}
\begin{figure}[h!]
    \centering
    \includegraphics[width=0.75\linewidth]{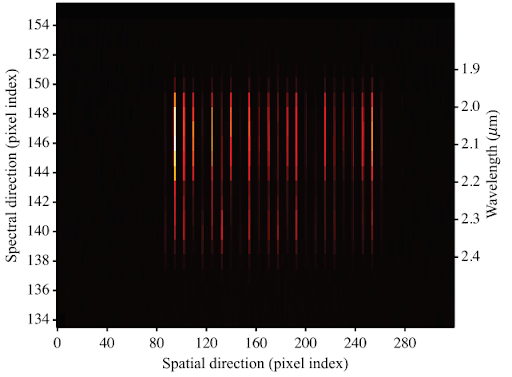}
    \caption{Example of MYSTIC ABCD observations as seen on the detector. Four beams are combined and output as 24 parallel lines, sampling four phases of the visibility contrast at each wavelength channel for 6 baselines\cite{mystic}}
    \label{fig:abcd}
\end{figure}

When it was first commissioned in 2021, MYSTIC only supported one observing mode, known as All-in-One (AIO), which is the six-telescope image plane combiner with photometric channels based on the MIRC-X design. Later in the summer of 2022, a cross-talk-resistant 4-telescope pairwise combiner (ABCD) was added to MYSTIC using the spare integrated optics (IO) chip from the VLTI/GRAVITY combiner. The ABCD mode is fully operational and is supported by the MIRC-X/MYSTIC observing software ecosystem. All of MYSTIC's spectral modes are available for use with the ABCD combiner. So far, it has been used to observe over 20 T Tauri stars down to $\sim 7.5$ Kmag. An example of the ABCD output beams can be seen in \autoref{fig:abcd}. Unfortunately, the data reduction pipeline for this mode is still under development, so no definitive conclusions can be made about the supposed increased sensitivity or calibration precision yet. 

\subsubsection{MIRC-X ABCD Mode with SPICA-FT}
\begin{figure}[h!]
    \centering
    \includegraphics[width=0.8\linewidth]{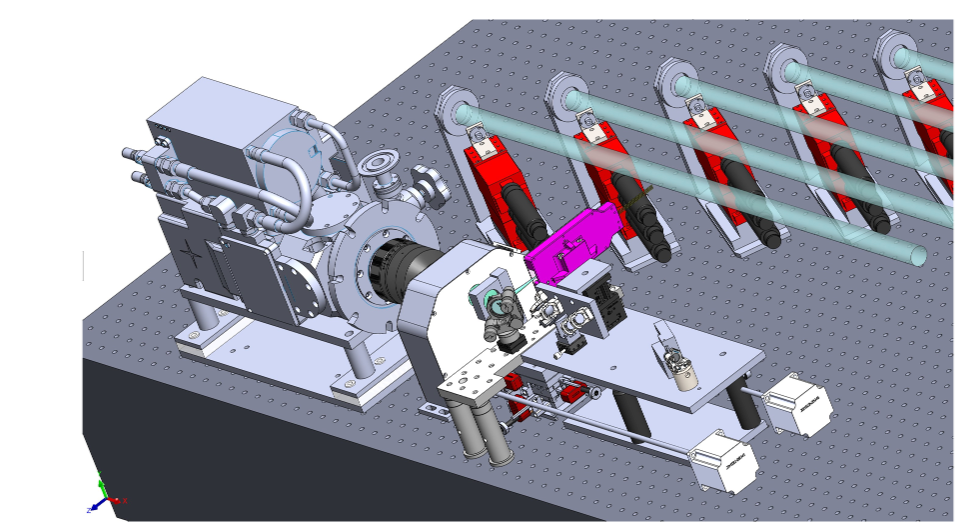}
    \caption{CAD view of the SPICA-FT (in pink) positioned near the MIRC-X spectrograph \cite{spicaft}}
    \label{fig:SPICAFT}
\end{figure}

In collaboration with the SPICA instrument (visible 6-beam combiner) team at the CHARA Array, MIRC-X has recently undergone some upgrades to integrate it with the new fringe tracker, SPICA-FT \cite{spicaft}. SPICA-FT is an H-band instrument that combines the light for six telescopes using a pairwise ABCD-encoded IO chip. The SPICA-FT IO combiner sits at the MIRC-X table and is located at the entrance of the instrument's spectrograph. A computer-aided design (CAD) of the combiner setup can be seen in \autoref{fig:SPICAFT}
\\

\begin{figure}[h!]
    \centering
    \includegraphics[width=0.85\linewidth]{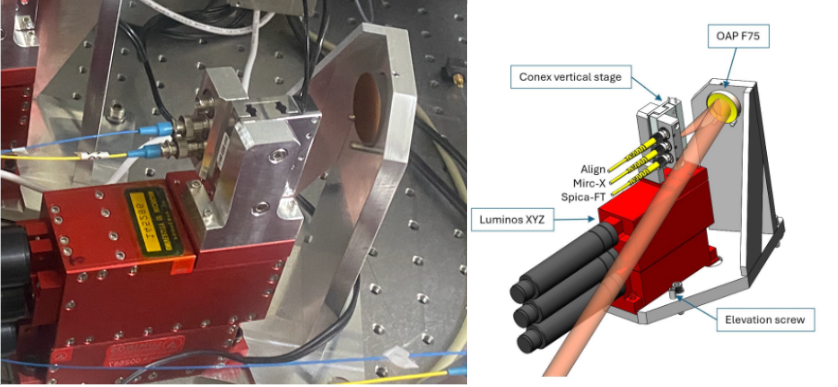}
    \caption{(Left) One of the six new fiber injection systems. The new OAP can be seen facing the new fiber switchyard that is holding the MIRC-X and SPICA-FT fibers (Right) Diagram showing the new fiber injection system}
    \label{fig:fiberinj}
\end{figure}

\begin{figure}[h!]
    \centering
    \includegraphics[width=\linewidth]{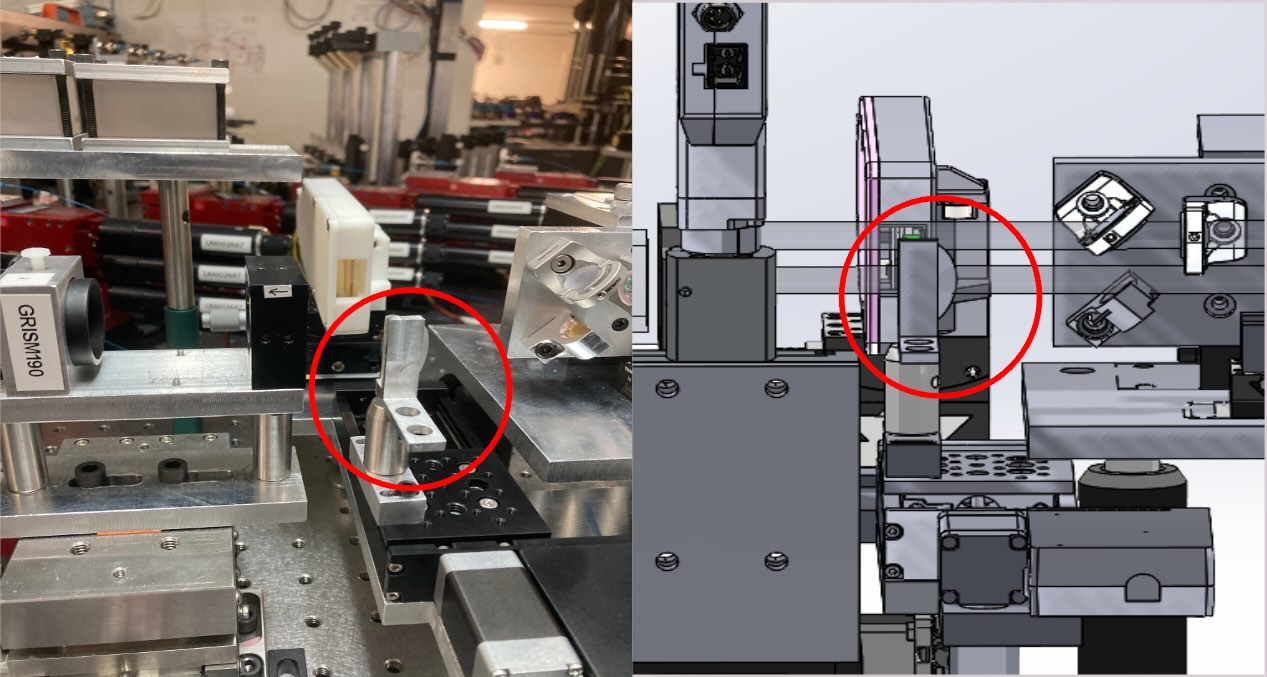}
    \caption{(left) The mirror switchyard, in the red circle, was successfully installed and integrated with the SPICA FT at the CHARA Array. (Right) CAD of the mirror switchyard}
    \label{fig:mirror}
\end{figure}

In order to feed different fibers with the incoming beam on the MIRC-X table, a new fiber injection system had to be designed. 
Each collimated beam of light is focused on the tip of the fiber by an off-axis parabolic (OAP) mirror. MIRC was designed with $f=60 mm$ OAPs, which was not an optimal focal length for SPICA-FT and MIRC-X. As a result, all six OAPs in MIRC-X were replaced with $f=75mm$ mm ones to provide better focus on the fiber. A motorized Newport CONEX translation stage was installed on top of the existing Luminos positioning mounts to avoid switching between the MIRC-X and SPICA-FT fibers by hand. This new fiber switchyard holds both fibers and enables the user to position the desired fiber at the OAP focal point, thus allowing for automated switching between combiners. An example of the new fiber injection system and a labeled diagram can be seen in \autoref{fig:fiberinj} \\
\newpage
Lastly, a moving pickoff mirror system was installed to feed the spectrograph from the user's choice of combiner. The pickoff mirror is circled in red in \autoref{fig:mirror}.
All of these upgrades have been successfully implemented and the accompanying operating software has been developed and is undergoing final commissioning. However, as for MYSTIC ABCD, the data reduction pipeline is still under development.

\subsubsection{J+H-band 6T Mode}
MIRC-X was designed as an H- and J-band combiner, although most users currently operate it solely in the H-band. The J-band can presently be used with only four telescopes due to magnification issues causing undersampling of some fringes. In addition, simultaneous H+J-band observing is unavailable because of excess dispersion and birefringence. A study was performed to calculate the dispersion effects between each pair of fibers and it showed that the greater the length difference between the pairs, the greater the dispersion effects \cite{aaron}. We report these calculations in \autoref{table:fiber}. 

\begin{table}[h!]

\centering

\begin{tabular}{| l | l | l | l |}

\hline
Beam & Length Difference & $\sim$Dispersion at best point & New fibers length difference \\
\hline
12 & 0.73 mm & 3.5 $\mu$m & 0.18 mm \\
\hline
23 & -0.64 mm & 3.0 $\mu$m & -0.63 mm \\
\hline
34 & -0.12 mm & 0.5 $\mu$m & -0.25 mm \\
\hline
45 & -1.45 mm & 5.0 $\mu$m & 0.44 mm \\
\hline
56 & 2.38 mm & -9.0 $\mu$m & -0.53 mm \\
\hline

\end{tabular}
\vspace{5pt}
\caption{Calculations by A. Labdon showing the length difference between each fiber pair and the corresponding dispersion. New fibers with better-matched lengths were bought and are pending testing }
\label{table:fiber}
\end{table}

Consequently, six new fibers that are better matched in length have been bought, and we report their length differences in the last column of \autoref{table:fiber}. The fibers have been assembled in a V groove and are pending testing at the University of Michigan (UM) lab. If the new fibers reduce the dispersion issues as expected, then we will install them in the combiner, allowing for simultaneous J+H observing. Testing of the new fibers is planned for this summer.

As mentioned, the J-band observing mode cannot operate in 6T due to a magnification issue. To address that, we plan to replace the current $f=100 mm$ collimator lens with a $f=75 mm$ one. The shorter focal length will allow for better sampling of the fringes in the shorter wavelength regime. Sadly, it has been difficult to source a suitable off-the-shelf $f=75 mm$ lens so the current plan is to design one in-house. The exact timeline for this upgrade is to be determined. The lens can be easily added to the system once it is available. 

\subsubsection{MIRC-X polarization mode}
A thorough discussion of the hardware and software upgrade of the MIRC-X polarization mode can be found in a previous SPIE proceeding by B. Setterholm\cite{pol}. The new mode is in the science verification stage now, which our team will be conducting over the summer.
\subsection{Automation}
\begin{figure}[h!]
    \centering
    \includegraphics[width=0.75\linewidth]{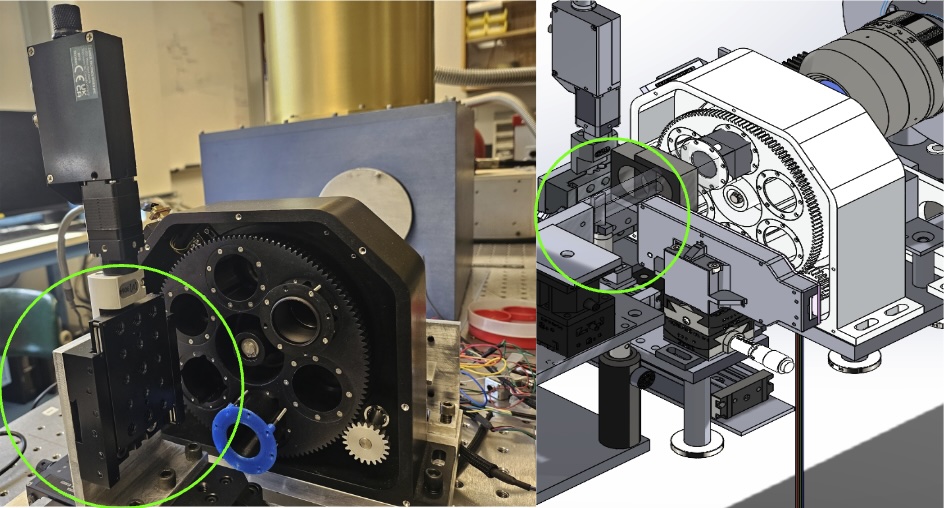}
    \caption{(left) The filter wheel is assembled and is being tested at UM lab. Special grating holders were 3D printed and can be seen in the blue. A translation stage (circled in green) will hold the different lenses and move desired on into the optical axis so observers can switch between modes (Right) CAD of the filter wheel, the grating holder, and the moving lens system.
}
    \label{fig:wheel}
\end{figure}

A significant portion of the upcoming upgrades will revolve around automating many aspects of the observing experience to reduce risk and optimize efficiency. 
First, we plan to install a motorized filter wheel in MIRC-X (similar to the one MYSTIC has). Currently, to change spectral modes, a person has to enter the lab on the mountain and change the dispersive optics by hand. To minimize the risks introduced by human error, the dispersive optics will be placed in a motorized wheel, and the observer will be able to control it through a Graphic User Interface (GUI). The filter wheel was designed by Universal Cryogenics and assembled at the UM lab. It is currently undergoing tests at UM to verify the accuracy and reproducibility of motion. Special holders had to be designed and 3D printed for the gratings to fit properly in the wheel slots. The new filter wheel and the 3D-printed grating holder are shown \autoref{fig:wheel}.

Second, our team has designed and built a lens translation module, which can be seen circled in green in \autoref{fig:wheel}. The moving lens system will be located in front of the filter wheel and will hold the $f=100 mm$ collimator lens as well as the $f=75 mm$ lens when it is available. This system is currently being tested at UM.

\begin{figure}
    \centering
    \includegraphics[width=0.75\linewidth]{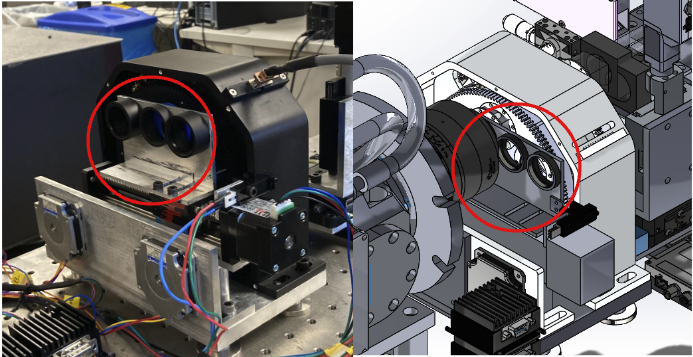}
    \caption{(Left) A bandpass slider will include a metrology laser blocker, and J and H filters  (Right) CAD of the bandpass filter slider}
    \label{fig:bandpass}
\end{figure}

Third, our team designed and built a bandpass filter slider, seen circled in red in \autoref{fig:bandpass}, that will hold the J and H filters as well as a laser line blocker filter. Users will be able to use a GUI to move the slider to the desired filter without having to enter the lab. Similar to the previous two upgrades, this system is being tested at the UM lab.

In July of this year, our team plans to transport the filter wheel, the moving lens system, and the filter slider to the CHARA array to be installed in MIRC-X. After this upgrade, MIRC-X will be fully automated, eliminating the need for humans to handle the optics by hand during day-to-day observing needs. 

\subsection{New Spectral Modes}
The MIRC-X filter wheels will introduce new spectral modes, detailed in \autoref{table:spec}. We plan to add new Volume Phase Holographic Gratings (VPHGs, R$\sim$4000 and 6000, see SPIE contribution in these proceedings) to both MIRC-X and MYSTIC to allow for spectral line experiments. There is an available slot in the MYSTIC filter wheel which will most likely accommodate the VPHG. The current available spectral modes in MYSTIC can be seen in \autoref{table:specK}.

\begin{table}[h!]
\parbox{.5\linewidth}{
\centering

\begin{tabular}{| l | l |}
\hline
Wheel 1 & Wheel 2 \\
\hline
empty & empty \\
R22 & \bf{Wollaston} \\
R50 & \bf{R182} \\
R102 & \bf{R625} \\
\bf{R2314} & R1170 \\
\bf{(VPHG)*} & \bf{(R300)}* \\
\hline
\end{tabular}
\vspace{5pt}
\captionsetup{justification=centering}
\caption{MIRC-X filter wheel elements.\\ New elements are denoted in bold font.\\**Available in 2025}
\label{table:spec}
}
\parbox{.5\linewidth}{
\centering

\begin{tabular}{| l | l |}
\hline
Wheel 1 & Wheel 2 \\
\hline
empty & empty \\
cold plug & R20 \\
R 278 & R49 \\
R49 & empty \\
narrow band & R981 \\
Wollaston & R1724 \\
\hline

\end{tabular}
\vspace{5pt}
\caption{MYSTIC filter wheel elements}
\vspace{25pt}
\label{table:specK}
}
\end{table}

\subsection{Improving Data Quality}

In the final section of instrument upgrades, we outline plans to enhance overall data quality of MIRC-X, which will involve constructing a new combiner and upgrading its C-RED ONE camera.

\begin{figure}[h!]
    \centering
    \includegraphics[width=\linewidth]{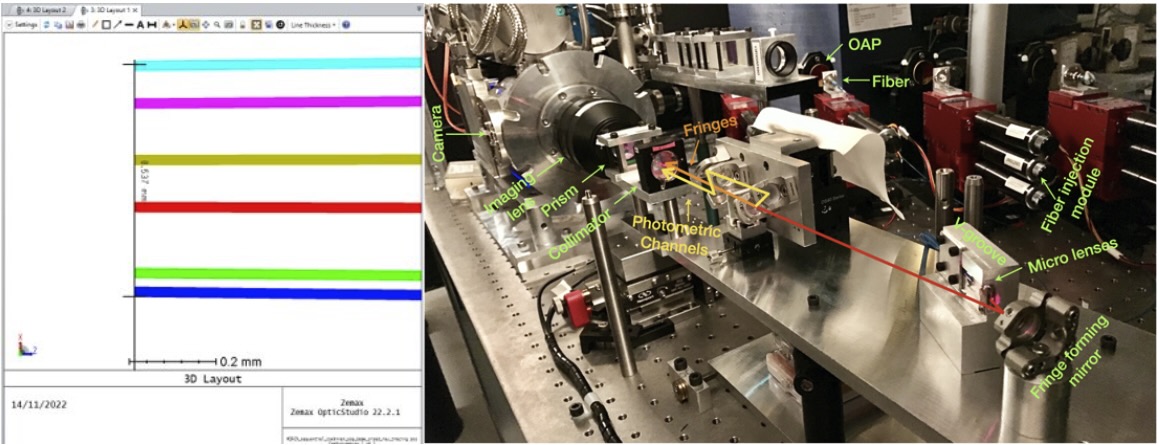}
    \caption{(Left) Zemax simulation of the beams at the current position of the focal point. The ray optics are not overlapping as shown. (Right) Current setup of the combiner. The colored lines show the path of the beams \cite{mircx1}  }
    \label{fig:oldaoi}
\end{figure}

In the MIRC-X combiner, the angle of incidence between the fringe-forming mirror and the microlens array is $22.5^\circ$. It was shown that this angle is too large and it leads to the beams overlapping 20 mm away from the focal point as can be seen in \autoref{fig:oldaoi}. This causes a minor reduction in the interference contrast. The new combiner will employ a shallower angle of incidence $\sim$ $2^\circ$ allowing the fringes to overlap optimally.

\begin{figure}[h!]
    \centering
    \includegraphics[width=\linewidth]{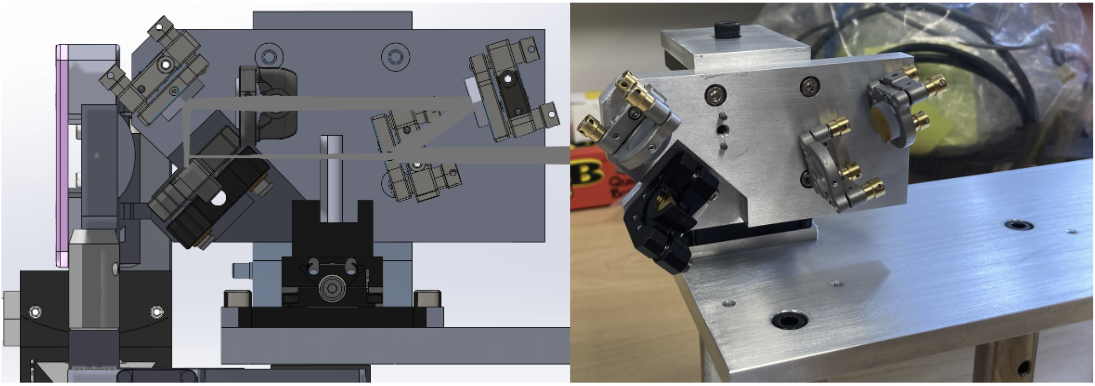}
    \caption{(Left) CAD of the new combiner showing the new adjustable mounts and the light path through the combiner. (Right) The new MIRC-X combiner will have adjustable mounts for select optics to allow for finer alignment. The combiner is being built at the UM lab.}
    \label{fig:newcombine}
\end{figure}

The new combiner will also have a new mounting plate for the optics, where optics will be situated in more adjustable mounts. The current optics are stationary and close inspection revealed that the beam was not centered on some of them. An off-center beam can lead to aberrations which will impact the image quality. By adding tip/tilt mounts to select optics, we increase the alignment degrees of freedom to allow for finer adjustments and minimize aberrations. The new combiner is being assembled at the UM lab and can be seen in \autoref{fig:newcombine} alongside a CAD of the final design. Most parts have either already been fabricated or are awaiting shipping from a manufacturer. Once all the parts arrive, the new combiner will be tested to verify that the planned updates do improve the signal and image quality in practice. The timeline for this upgrade is still to be determined pending lab testing.

\begin{figure}[h!]
    \centering
    \includegraphics[width=0.75\linewidth]{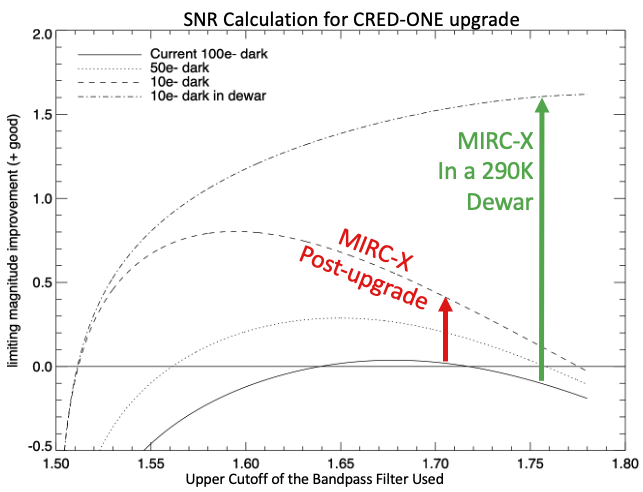}
    \caption{Calculation of the signal-to-noise ratio of the C-RED ONE camera as a function of the bandpass filter upper cutoff. The upgrade is expected to produce 10x lower dark current which should boost the sensitivity by $\sim 0.3$ magnitude. Placing MIRC-X in a 290 K dewar would boost the sensitivity even more, especially for the longer wavelengths.}
    \label{fig:snr}
\end{figure}

Lastly, Prof. Kraus has also funded an upgrade to the current CRED-ONE for MIRC-X with 10 times lower internal dark current, which could improve limiting magnitude by as much as $\sim 0.3$. Placing the combiner inside a dewar one day would add another magnitude boost. A calculation showing the SNR improvement we expect with this upgrade can be seen in \autoref{fig:snr} The camera improvements will also include less focal plane vibrations, better long-wave suppression, and removal of the cryocooler sine wave on signal. The camera upgrade is planned for January 2025 during the CHARA Array's winter shutoff. 
\newpage
\subsection{Summary}
We summarize the instrument upgrades discussed in this paper in \autoref{table:summary}. We list the main capabilities added to the combiners in the second column. The commissioning date or planned upgrade date is in the third column. These upgrades are ordered chronologically.
\begin{table}[h!]
\centering

\begin{tabular}{| p{2.5cm} | p{11cm} | p{2.1cm} |}
\hline
\textbf{Upgrade} & \textbf{Added Capabilities} & \textbf{Timeline} \\
\hline

MYSTIC ABCD mode & The new 4T ABCD IO chip adds a new sensitive cross-talk resistant mode to MYSTIC with better visibility amplitude calibration & Commissioned summer 2022 \\
\hline
SPICA-FT IO & Adds a new 6T ABCD observing mode fat H-band & Commissioned summer 2022 \\
\hline
Mirror Switchyard & Created a mirror switchyard to allow more instruments like SPICA FT to use the MIRC-X detector & Commissioned summer 2023 \\
\hline
New OAPs & Replaced the old f60 mm MIRC OAPs with f75 mm ones for a more suitable focal length for both MIRC-X and SPICA-FT & Commissioned  February 2024 \\
\hline
New Fiber Injection System & The new fiber switchyard allows users to smoothly switch between MIRC-X and SPICA-FT IO fibers & Commissioned February 2024 \\
\hline
2 Filter Wheels & Will install motorized filter wheels to enhance observing efficiency &  July 2024 \\
\hline
Lens Movement & A new lens movement system will accommodate multiple lenses including a new 75 mm lens to allow 6T J-band observing &  July 2024 \\
\hline
Upgraded C-RED ONE camera & The camera will reduce the dark current by tenfold, going from 100 e-/s to 10 e-/s. The higher SNR will enhance the sensitivity by $\sim$0.3 mag &  January 2025 \\
\hline
Adjustable Optics \& Shallow Angle of Incidence  & Will add adjustable mounts to the combiner optics to improve alignment and reduce aberrations. Reducing the angle between the microlens array and the fringe-forming mirror will decrease beam cross-talk & TBD \\
\hline
New Fibers  & Will replace the current fibers with ones that are better matched in length to reduce dispersion and allow for simultaneous J+H-band observations & TBD \\
\hline

\end{tabular}
\caption{A summary of commissioned and planned instrument upgrades and their timelines}
\label{table:summary}
\end{table}

\newpage
\acknowledgments 
 
SK acknowledges funding for MIRC-X received funding from the European Research Council (ERC) under the European Union's Horizon 2020 research and innovation programme (Starting Grant No. 639889 and Consolidated Grant No. 101003096). JDM acknowledges funding for the development of MIRC-X capabilities (NASA-XRP NNX16AD43G, NSF-AST 1909165) and MYSTIC (NSF-ATI 1506540, NSF-AST 1909165). We received funding from the European Union’s Horizon 2020 research and innovation programme under grant agreement No 101004719 (ORP).  S.K. also acknowledges support from an ERC Consolidator Grant (Grant Agreement ID 101003096) and STFC Consolidated Grant (ST/V000721/1). DM acknowledges funding from the European Research Council (ERC) under the European Union’s Horizon 2020 research and innovation programme (Grant agreement No. 101019653). Thanks to Samuel Mayberry from GSU shop for machining the mechanic associated with SPICA/FT

\bibliography{report} 

\begin{thebibliography}{10}

\bibitem{chara1}
{ten Brummelaar}, T.~A., {McAlister}, H.~A., {Ridgway}, S.~T., {Bagnuolo}, W.~G., J., {Turner}, N.~H., {Sturmann}, L., {Sturmann}, J., {Berger}, D.~H., {Ogden}, C.~E., {Cadman}, R., {Hartkopf}, W.~I., {Hopper}, C.~H., and {Shure}, M.~A., ``{First Results from the CHARA Array. II. A Description of the Instrument},'' {\em \apj}~{\bf 628},  453--465 (July 2005).

\bibitem{chara2}
{Gies}, D.~R., {Anderson}, M.~D., {Anugu}, N., {ten Brummelaar}, T.~A., {Castillo}, V., {Farrington}, C.~D., {Golden}, S., {Jones}, J.~W., {Klement}, R., {K{\"o}hler}, R., {Lanthermann}, C., {Ligon}, E.~R., {Majoinen}, O., {McAlister}, H.~A., {Ridgway}, S.~T., {Schaefer}, G.~H., {Scott}, N.~J., {Turner}, N.~H., {Vargas}, N.~L., {Webster}, L., and {Woods}, C., ``{Recent technical and scientific highlights from the CHARA Array},'' in [{\em Optical and Infrared Interferometry and Imaging VIII}{\nolinebreak\hspace{0.1em}]},  {M{\'e}rand}, A., {Sallum}, S., and {Sanchez-Bermudez}, J., eds., {\em Society of Photo-Optical Instrumentation Engineers (SPIE) Conference Series} {\bf 12183},  1218303 (Aug. 2022).

\bibitem{mirc1}
{Monnier}, J.~D., {Berger}, J.-P., {Millan-Gabet}, R., and {ten Brummelaar}, T.~A., ``{The Michigan Infrared Combiner (MIRC): IR imaging with the CHARA Array},'' in [{\em New Frontiers in Stellar Interferometry}{\nolinebreak\hspace{0.1em}]},  {Traub}, W.~A., ed., {\em Society of Photo-Optical Instrumentation Engineers (SPIE) Conference Series} {\bf 5491},  1370 (Oct. 2004).

\bibitem{mirc2}
{Monnier}, J.~D., {Pedretti}, E., {Thureau}, N., {Berger}, J.-P., {Millan-Gabet}, R., {ten Brummelaar}, T., {McAlister}, H., {Sturmann}, J., {Sturmann}, L., {Muirhead}, P., {Tannirkulam}, A., {Webster}, S., and {Zhao}, M., ``{Michigan Infrared Combiner (MIRC): commissioning results at the CHARA Array},'' in [{\em Advances in Stellar Interferometry}{\nolinebreak\hspace{0.1em}]},  {Monnier}, J.~D., {Sch{\"o}ller}, M., and {Danchi}, W.~C., eds., {\em Society of Photo-Optical Instrumentation Engineers (SPIE) Conference Series} {\bf 6268},  62681P (June 2006).

\bibitem{spots}
{Roettenbacher}, R.~M., {Monnier}, J.~D., {Korhonen}, H., {Aarnio}, A.~N., {Baron}, F., {Che}, X., {Harmon}, R.~O., {K{\H{o}}v{\'a}ri}, Z., {Kraus}, S., {Schaefer}, G.~H., {Torres}, G., {Zhao}, M., {Ten Brummelaar}, T.~A., {Sturmann}, J., and {Sturmann}, L., ``{No Sun-like dynamo on the active star $\zeta$ Andromedae from starspot asymmetry},'' {\em \nat}~{\bf 533},  217--220 (May 2016).

\bibitem{binary}
{Kloppenborg}, B., {Stencel}, R., {Monnier}, J.~D., {Schaefer}, G., {Zhao}, M., {Baron}, F., {McAlister}, H., {ten Brummelaar}, T., {Che}, X., {Farrington}, C., {Pedretti}, E., {Sallave-Goldfinger}, P.~J., {Sturmann}, J., {Sturmann}, L., {Thureau}, N., {Turner}, N., and {Carroll}, S.~M., ``{Infrared images of the transiting disk in the $\epsilon$ Aurigae system},'' {\em \nat}~{\bf 464},  870--872 (Apr. 2010).

\bibitem{nova}
{Schaefer}, G.~H., {Brummelaar}, T.~T., {Gies}, D.~R., {Farrington}, C.~D., {Kloppenborg}, B., {Chesneau}, O., {Monnier}, J.~D., {Ridgway}, S.~T., {Scott}, N., {Tallon-Bosc}, I., {McAlister}, H.~A., {Boyajian}, T., {Maestro}, V., {Mourard}, D., {Meilland}, A., {Nardetto}, N., {Stee}, P., {Sturmann}, J., {Vargas}, N., {Baron}, F., {Ireland}, M., {Baines}, E.~K., {Che}, X., {Jones}, J., {Richardson}, N.~D., {Roettenbacher}, R.~M., {Sturmann}, L., {Turner}, N.~H., {Tuthill}, P., {van Belle}, G., {von Braun}, K., {Zavala}, R.~T., {Banerjee}, D.~P.~K., {Ashok}, N.~M., {Joshi}, V., {Becker}, J., and {Muirhead}, P.~S., ``{The expanding fireball of Nova Delphini 2013},'' {\em \nat}~{\bf 515},  234--236 (Nov. 2014).

\bibitem{cred}
{Gach}, J.~L., {Feautrier}, P., {Stadler}, E., {Greffe}, T., {Clop}, F., {Lemarchand}, S., {Carmignani}, T., {Boutolleau}, D., and {Baker}, I., ``{C-RED one: ultra-high speed wavefront sensing in the infrared made possible},'' in [{\em Adaptive Optics Systems V}{\nolinebreak\hspace{0.1em}]},  {Marchetti}, E., {Close}, L.~M., and {V{\'e}ran}, J.-P., eds., {\em Society of Photo-Optical Instrumentation Engineers (SPIE) Conference Series} {\bf 9909},  990913 (July 2016).

\bibitem{cred2}
{Finger}, G., {Baker}, I., {Alvarez}, D., {Ives}, D., {Mehrgan}, L., {Meyer}, M., {Stegmeier}, J., and {Weller}, H.~J., ``{SAPHIRA detector for infrared wavefront sensing},'' in [{\em Adaptive Optics Systems IV}{\nolinebreak\hspace{0.1em}]},  {Marchetti}, E., {Close}, L.~M., and {Vran}, J.-P., eds., {\em Society of Photo-Optical Instrumentation Engineers (SPIE) Conference Series} {\bf 9148},  914817 (Aug. 2014).

\bibitem{mircx1}
{Anugu}, N., {Le Bouquin}, J.-B., {Monnier}, J.~D., {Kraus}, S., {Setterholm}, B.~R., {Labdon}, A., {Davies}, C.~L., {Lanthermann}, C., {Gardner}, T., {Ennis}, J., {Johnson}, K. J.~C., {Ten Brummelaar}, T., {Schaefer}, G., and {Sturmann}, J., ``{MIRC-X: A Highly Sensitive Six-telescope Interferometric Imager at the CHARA Array},'' {\em \aj}~{\bf 160},  158 (Oct. 2020).

\bibitem{mircx2}
{Anugu}, N., {Le Bouquin}, J.-B., {Monnier}, J.~D., {Kraus}, S., {Ennis}, J., {Lanthermann}, C., {Setterholm}, B.~R., {Davies}, C.~L., {ten Brummelaar}, T., {Haidar}, M., {Dubravec}, V., and {Peters}, S., ``{MIRC-X/CHARA: sensitivity improvements with an ultra-low noise SAPHIRA detector},'' in [{\em Optical and Infrared Interferometry and Imaging VI}{\nolinebreak\hspace{0.1em}]},  {Creech-Eakman}, M.~J., {Tuthill}, P.~G., and {M{\'e}rand}, A., eds., {\em Society of Photo-Optical Instrumentation Engineers (SPIE) Conference Series} {\bf 10701},  1070124 (July 2018).

\bibitem{yso1}
{Ibrahim}, N., {Monnier}, J.~D., {Kraus}, S., {Le Bouquin}, J.-B., {Anugu}, N., {Baron}, F., {Brummelaar}, T.~T., {Davies}, C.~L., {Ennis}, J., {Gardner}, T., {Labdon}, A., {Lanthermann}, C., {M{\'e}rand}, A., {Rich}, E., {Schaefer}, G.~H., and {Setterholm}, B.~R., ``{Imaging the Inner Astronomical Unit of the Herbig Be Star HD 190073},'' {\em \apj}~{\bf 947},  68 (Apr. 2023).

\bibitem{yso2}
{Kraus}, S., {Kreplin}, A., {Young}, A.~K., {Bate}, M.~R., {Monnier}, J.~D., {Harries}, T.~J., {Avenhaus}, H., {Kluska}, J., {Laws}, A. S.~E., {Rich}, E.~A., {Willson}, M., {Aarnio}, A.~N., {Adams}, F.~C., {Andrews}, S.~M., {Anugu}, N., {Bae}, J., {ten Brummelaar}, T., {Calvet}, N., {Cur{\'e}}, M., {Davies}, C.~L., {Ennis}, J., {Espaillat}, C., {Gardner}, T., {Hartmann}, L., {Hinkley}, S., {Labdon}, A., {Lanthermann}, C., {LeBouquin}, J.-B., {Schaefer}, G.~H., {Setterholm}, B.~R., {Wilner}, D., and {Zhu}, Z., ``{A triple-star system with a misaligned and warped circumstellar disk shaped by disk tearing},'' {\em Science}~{\bf 369},  1233--1238 (Sept. 2020).

\bibitem{yso3}
{Labdon}, A., {Kraus}, S., {Davies}, C.~L., {Kreplin}, A., {Zarrilli}, S., {Monnier}, J.~D., {Le Bouquin}, J.-B., {Anugu}, N., {Setterholm}, B., {Gardner}, T., {Ennis}, J., {Lanthermann}, C., {ten Brummelaar}, T., {Schaefer}, G., and {Harries}, T.~J., ``{Imaging the warped dusty disk wind environment of SU Aurigae with MIRC-X},'' {\em \aap}~{\bf 678},  A6 (Oct. 2023).

\bibitem{mystic}
{Setterholm}, B.~R., {Monnier}, J.~D., {Le Bouquin}, J.-B., {Anugu}, N., {Ennis}, J., {Jocou}, L., {Ibrahim}, N., {Kraus}, S., {Anderson}, M.~D., {Chhabra}, S., {Codron}, I., {Farrington}, C.~D., {Flores}, B., {Gardner}, T., {Gutierrez}, M., {Lanthermann}, C., {Majoinen}, O.~W., {Mortimer}, D.~J., {Schaefer}, G., {Scott}, N.~J., {ten Brummelaar}, T., and {Vargas}, N.~L., ``{MYSTIC: a high angular resolution K-band imager at CHARA},'' {\em Journal of Astronomical Telescopes, Instruments, and Systems}~{\bf 9},  025006 (Apr. 2023).

\bibitem{spicaft}
{Pannetier}, C., {B{\'e}rio}, P., {Mourard}, D., {Rousseau}, S., {Allouche}, F., {Dejonghe}, J., {Bailet}, C., {Lecron}, D., {Cassaing}, F., {Le Bouquin}, J.-B., {Perraut}, K., {Monnier}, J., {Anugu}, N., and {ten Brummelaar}, T., ``{SPICA-FT: the new fringe tracker of the CHARA array},'' in [{\em Optical and Infrared Interferometry and Imaging VIII}{\nolinebreak\hspace{0.1em}]},  {M{\'e}rand}, A., {Sallum}, S., and {Sanchez-Bermudez}, J., eds., {\em Society of Photo-Optical Instrumentation Engineers (SPIE) Conference Series} {\bf 12183},  1218309 (Aug. 2022).

\bibitem{aaron}
Labdon, A. Private communication.

\bibitem{pol}
{Setterholm}, B.~R., {Monnier}, J.~D., {Le Bouquin}, J.-B., {Anugu}, N., {Labdon}, A., {Ennis}, J., {Johnson}, K. J.~C., {Kraus}, S., and {ten Brummelaar}, T., ``{MIRC-X polarinterferometry at CHARA},'' in [{\em Optical and Infrared Interferometry and Imaging VII}{\nolinebreak\hspace{0.1em}]},  {Tuthill}, P.~G., {M{\'e}rand}, A., and {Sallum}, S., eds., {\em Society of Photo-Optical Instrumentation Engineers (SPIE) Conference Series} {\bf 11446},  114460R (Dec. 2020).

\end{thebibliography}
\bibliographystyle{spiebib} 

\end{document}